\documentclass[%
 reprint,
superscriptaddress,
 amsmath,amssymb,
 aps,
prl,
]{revtex4-1}
\usepackage{graphicx}
\usepackage{dcolumn}
\usepackage{bm}

\usepackage[utf8]{inputenc}
\usepackage[T1]{fontenc}
\usepackage{mathptmx} 
\DeclareMathAlphabet{\mathcal}{OMS}{cmsy}{m}{n} 
\usepackage{etoolbox}
\usepackage{braket}
\usepackage{lipsum}
\usepackage{xcolor}
\usepackage{hyperref}
\usepackage{amsmath}
\hypersetup{
    colorlinks=true,
    linkcolor=blue,
    citecolor=blue,
    filecolor=blue,
    urlcolor=blue,
    breaklinks=true  
}
\makeatletter
\def\@email#1#2{%
 \endgroup
 \patchcmd{\titleblock@produce}
  {\frontmatter@RRAPformat}
  {\frontmatter@RRAPformat{\produce@RRAP{*#1\href{mailto:#2}{#2}}}\frontmatter@RRAPformat}
  {}{}
}%
\makeatother

\newcommand\figscale{0.82}
\begin{document}

\preprint{AIP/123-QED}

\title{Angle-of-arrival detection of radio-frequency waves via Rydberg atom fluorescence imaging of standing waves in a glass vapor cell\\\vspace{.2cm}}
\author{Noah Schlossberger}
\email{noah.schlossberger@nist.gov}
\affiliation{National Institute of Standards and Technology, Boulder, Colorado 80305, USA}
\author{Rajavardhan Talashila}
\affiliation{Department of Electrical Engineering, University of Colorado, Boulder, Colorado 80309, USA}
\affiliation{National Institute of Standards and Technology, Boulder, Colorado 80305, USA}
\author{Nikunjkumar Prajapati}
\affiliation{National Institute of Standards and Technology, Boulder, Colorado 80305, USA}
\author{Christopher L. Holloway}
\affiliation{National Institute of Standards and Technology, Boulder, Colorado 80305, USA}

\date{\today}

\begin{abstract}
We present a method for measuring the angle-of-arrival of 40 GHz radio-frequency (RF) radiation by mapping the standing waves generated in a rectangular glass vapor cell. These standing waves have regular and well-defined structure from which we can infer the angle and sign of the wavevector of the RF field. We map the field using spatially resolved light sheet spectroscopy of Rydberg states of rubidium atoms in the cell. Unlike traditional phased arrays, this detection scheme is compact and low-complexity, has an active area of nearly 4$\pi$ steradians, and is sensitive to all RF polarizations. For in-plane measurements ($\phi = 0$), we demonstrate quantitative angle-of-arrival measurements with an uncertainty on the order of one degree in an 11~s measurement, and for out-of-plane measurements (arbitrary $\theta$,$\phi$), we demonstrate  angle-of-arrival detection with uncertainty on the order of several degrees.
\end{abstract}

\maketitle
\section{Introduction}
\vspace{-.4cm}
Angle-of-arrival measurements of radio-frequency (RF) radiation have many applications in industry and defense \cite{electronics11091321}, including wireless communication \cite{s24186144}, radar tracking \cite{skolnik2008radar}, electronic warfare \cite{989948}, and navigation systems \cite{evans1982advanced}. Classically, angle-of-arrival is determined using phased array antennas \cite{chen2010introduction}, which consist of spatially separated receivers that detect the relative phase at each point. While phased arrays are effective, they have several drawbacks: they are technologically complex, they scatter and absorb the incoming radiation, their active area is often limited to one hemisphere, and they tend to have a large radar cross-section. We develop a detection scheme that solves these issues by imaging the RF standing waves formed between the walls of a rectangular glass cell of atomic vapor.

Rydberg states of alkali atoms in atomic vapor have been demonstrated as effective traceable sensors of electromagnetic radiation with minimal perturbation to the fields they intend to measure  \cite{Schlossberger2024Nature, Sedlacek2012}. These sensors measure the electric field's perturbation to these states spectroscopically using electromagnetically induced transparency (EIT). A resonant RF field will induce a splitting in the energy of the Rydberg state proportional to the amplitude of the electric field. With a local oscillator, the phase of the incoming radiation can be measured \cite{SimonsPhase}, which allowed for angle-of-arrival detection in Rydberg sensors \cite{RobinsonAoA, gill2025microwavephasemappingangleofarrival, 10304615, gong2025rydbergatomicquantumreceivers, 9560547, Yan:23}, but at the cost  the local oscillator emitting radiation. To avoid this, we demonstrated an all-optical technique for measuring angle-of-arrival by generating a standing wave using a metal plate inside of the vapor cell and taking a two-point measurement of the resulting standing wave  \cite{talashila2025determininganglearrivalradio}. Limitations of this technique arise due to limited active angle and difficulty in determining the sign of the angle of arrival.

Recently, we developed a spatially resolved detection method using fluorescence imaging that allows the EIT spectrum to be measured simultaneously at every point in a plane, thus allowing us to ``image'' the RF field inside a vapor cell \cite{schlossberger2024twodimensionalimagingelectromagneticfields}. After using this technique to study surface effects \cite{patrick2025imaginginducedsurfacecharge}, we now turn it to angle-of-arrival measurements. It has been demonstrated that the dielectric walls of atomic vapor cells cause reflections and standing-waves of the RF incident fields inside  of the vapor cell \cite{imagingFan, imagingHolloway}. Here we use a rectangular vapor cell with thick walls of order 1/4 of the RF wavelength in order to generate strong standing patterns, which have well-defined structure that is dependent on the angle-of-incidence. We then image these standing waves using fluorescent detection, and fit the profiles to determine the angle-of-arrival. In this scheme, the interaction of the radiation with the vapor cell walls, which limits the accuracy of phase-based angle-of-arrival measurements \cite{GTRItheoryAoA}, is actually the interaction that gives angle-of-arrival resolution.
\vspace{-.55cm}
\section{Detection Scheme}
\vspace{-.4cm}
\begin{figure}[h]
\includegraphics[scale = \figscale]{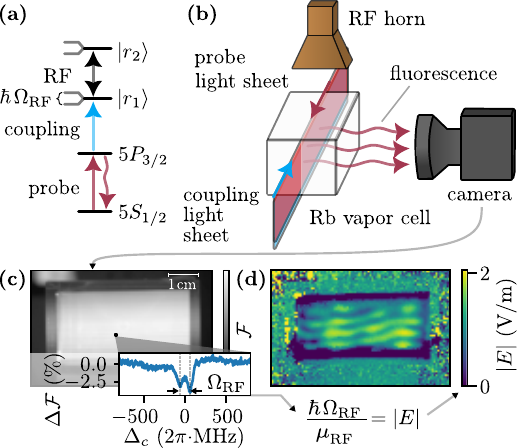}
\caption{Scheme for measuring the electric field in a plane using Rydberg atoms. (a) Energy level diagram of the spectroscopic readout. (b) Layout of the detection scheme. (c) The fluorescence $\mathcal{F}$ is recorded on the image as the detuning of $\Delta_c$ of the coupling laser is scanned. The profile of the change in fluorescence $\Delta \mathcal{F}$ as a function of the laser frequency is fit at each pixel to find the induced splitting $\Omega_{RF}$. (d) The result is a spatial profile of $|E|$ using Eq. \ref{eq:ATsplitting}.}
\label{fig:layout}
\end{figure}

Our detection scheme uses the same principle as Ref. \cite{schlossberger2024twodimensionalimagingelectromagneticfields}. We probe the Rydberg state via two-photon EIT in $^{85}$Rb (Fig. \ref{fig:layout}a). Here, $\ket{r_1}$ and $\ket{r_2}$ are two Rydberg states, with which the RF will be resonant. We read out the transparency via the decrease in fluorescence of the intermediate $5P_{3/2}$ state, which we image to give spatial resolution (Fig. \ref{fig:layout}b). Due to the Autler-Townes effect, the RF field induces an energy-splitting in $\ket{r_1}$ given by
\begin{equation}
    \hbar \Omega_\textrm{RF} = \mu_\textrm{RF}|E|, \label{eq:ATsplitting}
\end{equation}
where $\hbar$ is Planck's constant, $\Omega_\textrm{RF}$ is the Rabi frequency of the RF field, $\mu_{\textrm{RF}}$ is the transition dipole moment between states $\ket{r_1}$ and $\ket{r_2}$, and $|E|$ is the amplitude of the RF field applied. We read out this energy-splitting spectroscopically by scanning the frequency of the coupling laser (Fig. \ref{fig:layout}c). Fitting the spectrum measured at each pixel to find the Rabi frequency $\Omega_\textrm{RF}$, and thus the field magnitude via Eq. \ref{eq:ATsplitting}, we get a spatially resolved image of the RF field magnitude (Fig. \ref{fig:layout}d). One field distribution measurement takes approximately 11~s, limited by the integration time required for adequate signal-to-noise.
\vspace{-.4cm}
\section{\label{sec:level1}Planar measurements ($\phi = 0$)}
\begin{figure}[h]
\includegraphics[scale = \figscale]{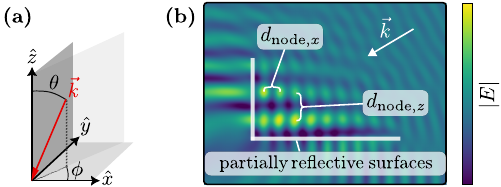}
\caption{Principle of the angle of arrival scheme. (a) Definition of the coordinate system used in this paper. (b) Modeled standing wave pattern of light incident on two orthogonal partially reflective surfaces. The field is presented in the $x$-$z$ plane with a $k$-vector at $\theta=60^\circ$, $\phi = 0$.}
\label{fig:2dscatteringtheory}
\end{figure}

\vspace{-.4cm}
Consider the case of wavevectors constrained to the $x$-$z$ plane, or in spherical coordinates, $\phi = 0$ (Fig. \ref{fig:2dscatteringtheory}a). In the presence of two partially reflective surfaces in the $z$-$y$ and $x$-$y$ planes, the reflected fields interfere with each other, creating the field distribution shown in Fig. \ref{fig:2dscatteringtheory}b.

\begin{figure*}[]
    \includegraphics[scale = \figscale]{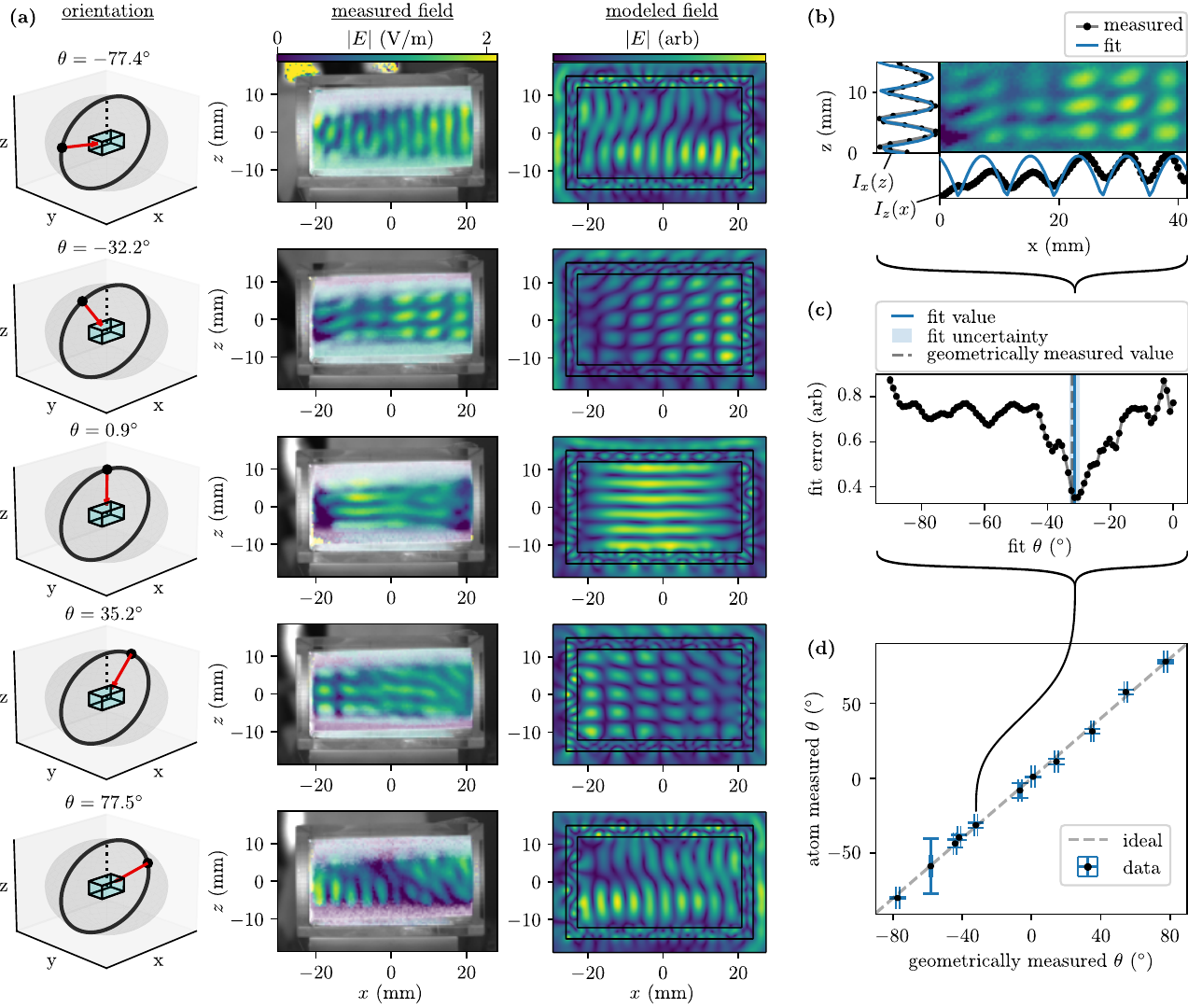}\\
\caption{Standing waves of 36.8951 GHz radiation measured via the 
$\ket{r_1} = 39D_{5/2} \rightarrow \ket{r_2}=40P_{3/2}$ level scheme at various values of $\theta$. (a) Left: the geometrically measured angle to the horn antenna. Middle: grayscale images of the vapor cell, with the measured field overlayed in color. The field is masked by the signal-to-noise ratio of the EIT spectrum such that the field is only shown where it is accurately measured. Right: Finite element simulation of the vapor cell in two dimensions assuming an incident plane wave. (b) At \mbox{$\theta = -32.2^\circ$} , the fluorescence image of the field is integrated in $x$ and $z$, and fit to Eq. \ref{eq:nodespacing} to find $\theta$. (c) The fit error defined in Eq. \ref{eq:fiterror} is plotted for this field distribution. (d) The fit is performed over a series of field distributions measured at different angles and the fitted atom measured $\theta$ is compared to the geometrically measured $\theta$ from the vapor cell to the RF horn.
}
\label{fig:phiequals0measurements}
\end{figure*}

The result is a standing wave whose amplitude is a regular grid of points with a well-defined spacing that is dependent on $\theta$. This can be approximated by summing the incident and reflected waves, and integrating over time to find the the root mean square (RMS) amplitude of the field. Doing this yields
\begin{align}
    E_\textrm{RMS} &= E_0 + E_1 |\cos(k_x x) \cos (k_z z)| 
    \\&= E_0 + E_1 \left|\cos\left(x \frac{2\pi}{\lambda} \sin\theta  \right) \cos\left(z \frac{2\pi}{\lambda} \cos\theta \right)\right|,
\end{align}
where $\vec{k}$ is the wavevector and $\lambda$ is the wavelength of the RF field. Here, $E_0$ is a constant field amplitude due to the non-interfering portion of the field and $E_1$ is the amplitude of the standing waves. From this we can get the node spacing
\begin{equation}
    \begin{cases}d_{\textrm{node},x} = \frac{\lambda}{2\sin\theta}\\
    d_{\textrm{node},z} = \frac{\lambda}{2\cos\theta}
    \end{cases}.
    \label{eq:nodespacing}
\end{equation}
Note that we get purely vertical fringes at $\theta = 0$ and only horizontal fringes at $\theta = \pi/2$.

In a rectangular vapor cell, the glass walls of the vapor cell are partially reflective to RF radiation, representing the set of partially reflective surfaces described above. The other two walls of the vapor cell also cause reflections, but this mostly acts to attenuate the field inside the cell and introduces phase distortions due to its finite thickness. The simple model still qualitatively predicts the standing waves observed inside the vapor cell and encapsulates the relevant degrees of freedom.

The standing waves are measured at $\phi = 0$ for a variety of $\theta$ in Fig. \ref{fig:phiequals0measurements}a. We compare the measured field distributions to a finite element simulation using the measured cell geometry. To make the simulation computationally tractable, we simulate in two dimensions (assuming the vapor cell is infinitely long in $y$). We get qualitative agreement between the measured and simulated fields. One feature of note is that the fields have complicated structure near the entry point, and have stronger and more regular standing wave patterns on the other side of the cell. This is most likely due to warping of the phasefront as the wave enters the cell due to refraction through the glass. By the time the wave reaches the other end of the cell, these near field effects are minimized since the wave has traveled several wavelengths, and the reflections off the far side do not exhibit this effect since they have not passed through any glass. While these distortions near the entry point may reduce fitting robustness, they can differentiate the entrance and exit ports, i.e. the \textit{sign} of the k-vector.

To quantitatively determine the angle-of-arrival, we crop the image to the region where the fields are well-resolved and integrate the measured field distributions along $x$ to produce a one-dimensional profile $I_x(z)$ and along $z$ to produce a one-dimensional profile along $I_z(x)$ (Fig. \ref{fig:phiequals0measurements}b). These profiles will then exhibit the periodicity given by Eq. \ref{eq:nodespacing}, and the integration will average away the complicated near-field structure at the location of incidence. As we know $\lambda$ and have calibrated the pixel size using the width of the vapor cell, we can then perform a combined fit of these two profiles to recover $\theta$. We minimize the residual of a simple squared sinusoid floating $\theta$ and an initial phase for each dimension $\phi_{0,x}$ and $\phi_{0,z}$. We define the fit error as:
\begin{multline}
\textrm{fit error}(\theta) \equiv \min_{\phi_{0,x}}\Big( \min_{\phi_{0,y}} \Big(
\\
\!\!\!\!\frac{1}{L_x}\int_0^{L_x} \left| \left|\cos\left(\frac{2\pi \sin \theta}{\lambda}x + \phi_{0,x}\right)\right| - I_z(x)\right| dx \\
+ 
\frac{1}{L_z}\int_0^{L_z} \left|\left|\cos\left(\frac{2\pi \cos \theta}{\lambda}z + \phi_{0,z}\right)\right| - I_x(z)\right| dz\Big) \Big), \label{eq:fiterror}
\end{multline}
where $L_x$ and $L_z$ are the length of the region of measured field in $x$ and $z$ respectively. The fit $\theta$ is found by minimizing this function, and the uncertainty of the fit is taken to be the width at which the minimum in fit error increases by five percent of the overall dip height (Fig. \ref{fig:phiequals0measurements}c). In Fig. \ref{fig:phiequals0measurements}d, we perform this fit over a range of angles limited by our optical setup. We find agreement between the geometrically measured angle between the vapor cell and the horn and the fit angle-of-arrival, with about one degree of uncertainty in the $\theta$ measured with the standing waves.

\vspace{-.4cm}
\section{3D Extension}
\vspace{-.4cm}
\begin{figure*}
\includegraphics[width = .8\textwidth]{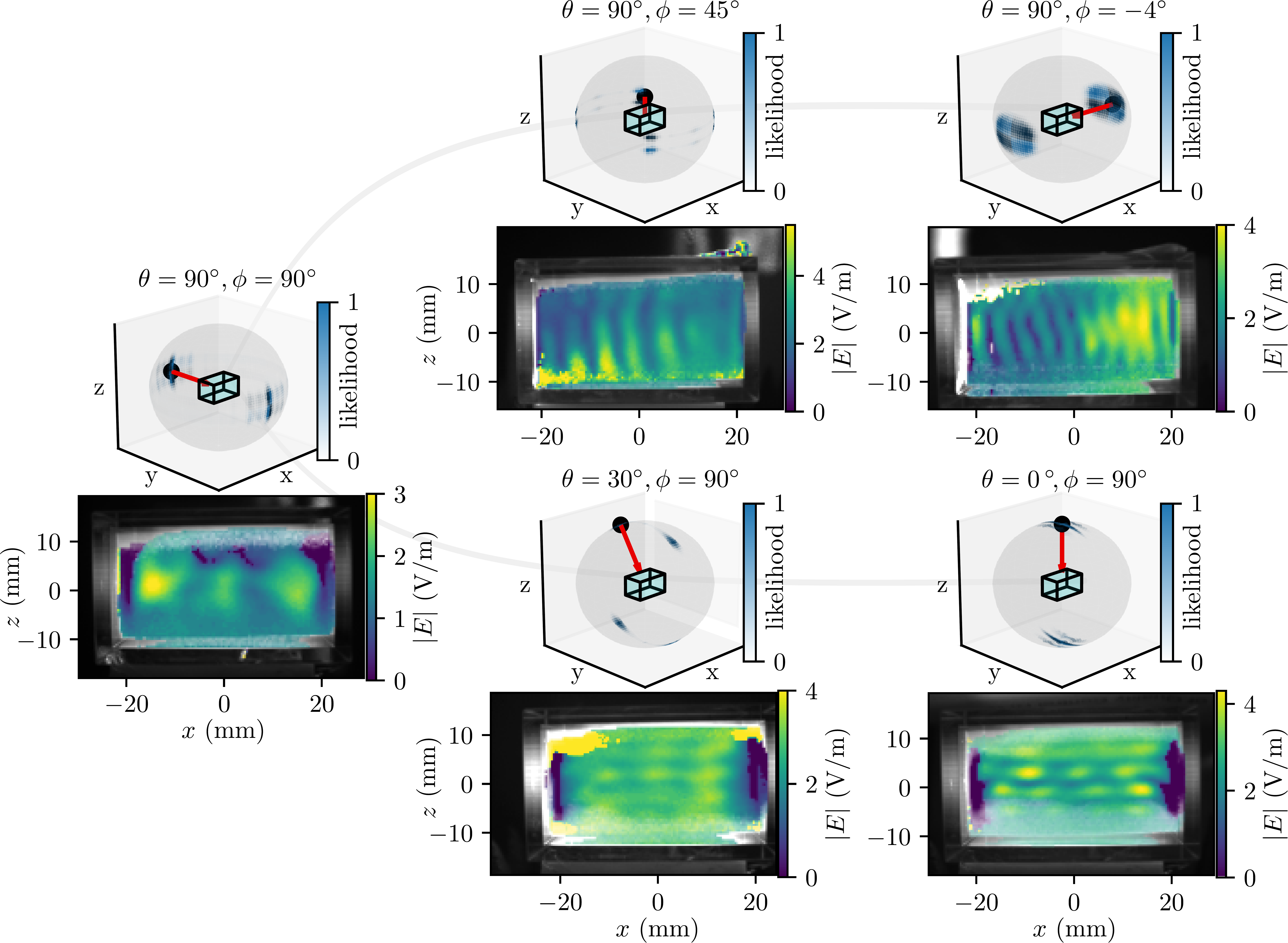}
\caption{Standing waves of 37.0727 GHz radiation measured via the $\ket{r_1} = 40D_{5/2} \rightarrow \ket{r_2} = 39F_{7/2}$ level scheme at various values of $\theta$ and $\phi$. Above each field distribution, the geometrically measured angle-of-arrival is indicated with a red arrow. We then calculate the likelihood estimator given by Eq. \ref{eq:likelihood} and shade the unit sphere to indicate the estimated three-dimensional angle-of-arrival. Unlike in the planar measurements, it is not clear that we can discern the sign of the angle-of-arrival, so all quadrants are shaded.}
\label{fig:phi_meas}
\end{figure*}

Let us now consider a general $k$-vector in three dimensions, allowing a nonzero $\phi$. The equivalent node spacing as in Eq. \ref{eq:nodespacing} becomes:
\begin{equation}
    \begin{cases}d_{\textrm{node},x} = \frac{\lambda}{2\sin\theta\cos\phi}\\
    d_{\textrm{node},y} = \frac{\lambda}{2\sin\theta\sin\phi}\\
    d_{\textrm{node},z} = \frac{\lambda}{2\cos\theta}\\
    \end{cases}
\end{equation}
Since we are measuring in the $x$-$z$ plane we do not have access to $d_{\textrm{node},y}$, but $d_{\textrm{node},x}$ and $d_{\textrm{node},z}$ contain enough degrees of freedom to constrain $\theta$ and $\phi$.

For a field incident from an arbitrary direction, the polarization of the radiation is most likely not aligned in a particular axis. The $D_{5/2}\rightarrow P_{3/2}$ transition used in the previous section is sensitive to the polarization, as the relative polarizations between the optical and RF photons can lead to population in $m_F = 5/2$ which does not have any dipole-allowed transitions for $\pi$-polarized microwaves and is thus insensitive to RF radiation. This leads to a central peak in the EIT spectra. While this can be used to extract information about the polarization  \cite{talashila2025determininganglearrivalradio}, it significantly reduces the signal-to-noise ratio of the split portion of the population for certain polarizations. To make a robust angle-of-arrival detector for an arbitrary polarization, we now move to a $D_{5/2} \rightarrow F_{7/2}$ transition for sensing so that there will be no RF-insensitive population. In this scheme, the polarization can affect the strength of the splitting (due to the different transition dipole moments between $m_F$ states), but this acts as an overall scaling factor of the standing wave profile, which still allows accurate angle-of-arrival estimation.

We perform the measurement at a variety of $\theta$ and $\phi$ values in Fig. \ref{fig:phi_meas}, demonstrating the $\cos\theta$ dependence in tilting axis and the $\cos\phi$ dependence in the other axis.  For each angle, we once again integrate both dimensions and perform a simultaneous fit to find the angle of incidence. Now, the objective function is given by

\begin{multline}
\textrm{fit error}(\theta, \phi) \equiv \min_{\phi_{0,x}}\Big( \min_{\phi_{0,y}} \Big(
\\
\;\;\;\;\;\;\;\;\frac{1}{L_x}\int_0^{L_x} \left|\left|\cos\left(\frac{2\pi \sin \theta\cos\phi}{\lambda}x + \phi_{0,x}\right)\right| - I_z(x)\right| dx \\
+ 
\frac{1}{L_z}\int_0^{L_z} \left|\left|\cos\left(\frac{2\pi \cos \theta}{\lambda}z + \phi_{0,z}\right)\right| - I_x(z)\right| dz\Big) \Big). \label{eq:fiterror3D}
\end{multline}
We then define a ``likelihood'' estimator as
\begin{equation}
\textrm{likelihood} (\theta, \phi) = \left(\frac{\max(\textrm{fit error}) - \textrm{fit error}(\theta, \phi)}{\max(\textrm{fit error}) - \min(\textrm{fit error})} \right)^N, \label{eq:likelihood}
\end{equation}
where $N$ is empirically chosen to have value 12 in order to give contrast.

With this, we are able to discern $|k_x|$, $|k_y|$, and $|k_z|$ with a single planar measurement. The angular uncertainty in the three-dimensional case varies for different angles-of-incidence, but is generally on the order of several degrees for both $\theta$ and $\phi$. In this configuration, we are subject to some error due to mis-alignment between the light sheets and the $x-z$ plane of the vapor cell, as this can cause the spatial distribution to reflect the $z$-axis nodes. We believe this is responsible for the field ``hot spots'' that appear in our $\theta = 90^\circ$ measurements.
\vspace{-.6cm}
\section{Conclusion}
\vspace{-.2cm}
We have demonstrated angle-of-arrival detection of 40~GHz radiation by measuring the periodicity of the standing wave pattern in a rectangular vapor cell. In a plane, we quantitatively determine the angle with approximately one degree of uncertainty in an 11~s measurement. We measure over the range of our swingarm, from -80 to 80 degrees, but the technique has a capability to detect over a full 360 degrees. In addition, we demonstrate that for most angles we can determine the sign of the $k$-vector in addition to the angle. We generalize this to three dimensional angle-of-arrival detection, showing that the standing waves are sensitive to both $\theta$ and $\phi$.

In the future, developing a fiber-coupled version of this probe would improve the accuracy of the angle-of-arrival detector by reducing the scattering of the RF field induced by nearby optics, and would allow one to realize a true $4\pi$ steradian active solid angle, which is currently limited by our optical table.
\vspace{-.9cm}
\section{Methods}
\vspace{-.4cm}
Both beams were expanded into light sheets using prism pairs and cylindrical lens pairs. They were over-expanded and clipped to 25~mm in order to produce a flat-top beam profile along $z$. Both beams had a Gaussian profile in $y$, with a full width at half maximum (FWHM) of 1.00~mm. Both light sheets were polarized along the $y$-axis. The coupling and probe laser powers were 209~mW and 2.9~mW respectively. The fluorescence was imaged through a 715~nm colored glass longpass filter by a complementary metal oxide semiconductor (CCD) camera with a commercial compound lens placed 9~cm away from the plane of excitation. The probe laser was locked to a reference vapor cell using saturated absorption spectroscopy. The coupling laser was scanned over a range of $2\pi \cdot 244$~MHz in 11~s while the camera recorded at a frame rate of 20 frames per second (an integration time of 50~ms per frame). The RF horn was mounted on a rotatable swing-arm with its center axis coincident with the center of the cell. The measured $\theta$ was found in software from a webcam image of the swing-arm taken normal to the plane of rotation. A reference edge was placed parallel to the wall of the cell to define $\hat z$, and the angle of the swing-arm was referenced to this surface. The uncertainty in this measured angle-of-arrival was around one degree,  arising from geometric optics and image projection effects. The distance from the horn to the cell was 80~cm, with a horn aperture of 10$\times$7~mm, justifying the treatment of the incident field as a plane wave. The vapor cell had outer dimensions of 30~mm in $x$, 30~mm in $y$, and 49.5~mm in $z$. The walls, which were 3~mm thick, were made from PYREX\textsuperscript{\textregistered}
 glass \cite{NISTDisclaimer} which has a relative permittivity of 4.6 at 1~MHz \cite{corning_sg3_2023}. We used this value in the finite element model in Fig. \ref{fig:phiequals0measurements}, assuming (for lack of sufficient data) that the dielectric constant is similar near 40~GHz.
\vspace{-.6cm}
\section*{Acknowledgments}
\vspace{-.4cm}
This research  was supported by NIST under the NIST-on-a-Chip program.  A contribution of the U.S. government, this work is not subject to copyright in the United States. 
\vspace{-.6cm}
\section*{Data availability}
\vspace{-.4cm}
All of the data presented in this paper and used to support the conclusions of this article is published under the identifier doi:10.18434/mds2-3102.
\vspace{-.6cm}
\section*{Conflicts of Interest}
\vspace{-.4cm}
The authors declare no conflict of interest.
\interlinepenalty=10000

\begin{thebibliography}{23}
\bibitem [{NIS()}]{NISTDisclaimer}%
  \BibitemOpen
  \href@noop {} {}\bibinfo {note} {Certain equipment, instruments, software, or materials are identified in this paper in order to specify the experimental procedure adequately. Such identification is not intended to imply recommendation or endorsement of any product or service by NIST, nor is it intended to imply that the materials or equipment identified are necessarily the best available for the purpose.}\BibitemShut {Stop}%
\end{thebibliography}
%

\end{document}